\documentclass{IEEEtaes}

\usepackage{color,array,amsthm}
\usepackage{graphicx}
\usepackage{amsfonts,amssymb}
\usepackage{graphicx}
\usepackage{subcaption}

\jvol{XX}
\jnum{XX}
\jmonth{XXXXX}
\paper{1234567}
\pubyear{2025}
\doiinfo{TAES.2025.Doi Number}

\begin{document}

\title{An Accelerated Camera 3DMA Framework for Efficient Urban GNSS Multipath Estimation}

\author{SHIYAO LV}

\author{XIN ZHANG}

\author{XINGQUN ZHAN}
\affil{Shanghai Jiao Tong University, Shanghai, China}


\receiveddate{Manuscript received XXXXX 00, 0000; revised XXXXX 00, 0000; accepted XXXXX 00, 0000.\\
This work was supported by the National Key R$\&$D Program of China (2022YFB3904401). }

\corresp{{\itshape  (Corresponding author: Xin Zhang.)}}

\authoraddress{Shiyao Lv, Xin Zhang and Xingqun Zhan are with the School of
 Aeronautics and Astronautics, Shanghai Jiao Tong University, Shanghai
 200240, China, E-mail: (lvshiyao@sjtu.edu.cn;xin.zhang@sjtu.edu.cn; xqzhan@sjtu.edu.cn).}

\editor{ }
\supplementary{ }

\markboth{LV ET AL.}{ACCELERATED 3DMA FRAMEWORK FOR GNSS MULTIPATH}
\maketitle

\begin{abstract}Robust Global Navigation Satellite System (GNSS) positioning in urban environments is still plagued by multipath effects, particularly due to the complex signal propagation induced by ubiquitous surfaces with varied radio frequency reflectivities. Current 3D Mapping Aided (3DMA) GNSS techniques show great potentials in mitigating multipath but face a critical trade-off between computational efficiency and modeling accuracy. Most approaches often rely on offline outdated or oversimplified 3D maps, while real-time LiDAR-based reconstruction boasts high accuracy, it is problematic in low laser reflectivity conditions; camera 3DMA is a good candidate to balance accuracy and efficiency but current methods suffer from extremely low reconstruction speed, a far cry from real-time multipath-mitigated navigation. This paper proposes an accelerated framework incorporating camera multi-view stereo (MVS) reconstruction and ray tracing. By hypothesizing on surface textures, an orthogonal visual feature fusion framework is proposed, which robustly addresses both texture-rich and texture-poor surfaces, lifting off the reflectivity challenges in visual reconstruction. A polygonal surface modeling scheme is further integrated to accurately delineate complex building boundaries,  enhancing the reconstruction granularity. To avoid excessively accurate reconstruction, reprojected point cloud multi-plane fitting and two complexity control strategies are proposed, thus improving upon multipath estimation speed. Experiments were conducted in Lujiazui, Shanghai, a typical multipath-prone district. The results show that the method achieves an average reconstruction accuracy of 2.4 meters in dense urban environments featuring glass curtain wall structures, a traditionally tough case for reconstruction, and achieves a ray-tracing-based multipath correction rate of 30 image frames per second, 10 times faster than the contemporary benchmarks.
\end{abstract}


\section{INTRODUCTION}
M{\scshape ultipath-resistant}  Global Navigation Satellite System (GNSS) positioning is critical for land or low altitude airspace autonomy in dense urban environments such as Lujiazui, Shanghai or Canary Wharf, London. The prevalence of signal reflecting surfaces introduces severe multipath inducing positioning errors of more than 10 meters in dense urban canyons \cite{new1},\cite{bib2}, a lot more than that of a multipath-free case. 

Recent studies have highlighted the limitations of conventional mitigation techniques. At different stages of signal processing chain, various methods have been developed. Dual-polarized antennas can reduce multipath errors by 30-50 $\%$ under open-sky conditions, but fail in urban canyons due to polarization distortion caused by multi-bounce reflections \cite{bib3}. Receiver correlator techniques, while effective for medium-range multipath, struggle with sub-meter short-range multipath delays \cite{bib4}. Correlator techniques such as the Extended Double-Delta Correlator improves suppression of short-delay multipath by enhancing correlation symmetry and delay resolution \cite{new4} but it has limited inhibition ability against long-delay multipaths in dense urban scenarios. Vector tracking methods face the risk of error contamination in multichannel GNSS receivers, possibly degrading positioning accuracy by 20-40 $\%$ under dynamic urban conditions \cite{bib5}. Yao et al. \cite{bib6} improved upon sky segmentation with fisheye cameras to achieve a 70 $\%$ non-line-of-sight (NLOS) rejection rate. However, the rejection rate can hardly be improved due to its inability to track each and every reflected signal ray since it was not based on signal propagation path reconstruction. Therefore, ray-tracing is the most promising method because of its ability to accurately model signal propagation paths in complex environments, including multiple reflections, diffractions and other signal-surface interaction. Moreover, ray tracing methods are based on the physical principles of electromagnetic wave propagation, do not rely on empirical assumptions, and are suitable for dynamic and unknown environments. However, three challenges exist for an effective ray tracing-based multipath mitigation: two for reflective surface reconstruction and one for multipath delay estimation based on the reconstructed results.

The first challenge is failure to reconstruct surfaces with poor texture, specifically buildings with glass curtain walls, which are prone to signal reflection but pose great difficulties for feature extraction by LiDARs \cite{bib7} or cameras \cite{bib6}. In addition, reflections of other buildings on the glass curtain wall will pose an even greater challenge for correct feature extraction and association \cite{bib8}. This challenge direcly affects robustness and accuracy of ray tracing. Wen et al. \cite{bib7} tried LiDAR-assisted NLOS detection through real-time point cloud registration but LiDAR may fail to receive the return signal from surfaces with low reflectivity, limiting its reliability. Simulation modeling is usually based on existing 3D geographic information platforms (e.g. Google Earth, OpenStreetMap) \cite{new8} to build a virtual urban environment, and relies on ray tracing technology to simulate signal propagation paths. However, this method suffers from insufficient reconstruction accuracy, slow 3D map update rate and high computational complexity. In terms of algorithms for 3D reconstruction, traditional Structure from Motion (SfM) algorithms such as the combination of COLMAP \cite{bib9} and OPENMVS \cite{bib10} etc. have low reconstruction accuracy in the poor texture region of glass curtain wall reflection since its feature extractor’s incompetence in dealing with surfaces with ‘ghost’ images, i.e., neighboring building’s reflection or repetitive texture patterns. In recent years, deep learning techniques have been widely used in the reconstruction of complex scenes \cite{bib11},\cite{bib12}, but still face the bottleneck of missing textures and computational efficiency. Depth Anything \cite{bib13} and PSMNet \cite{bib14} are able to provide depth information which is pivotal in 3D reconstruction, in texture-rich conditions, but for poor texture or when subject to lighting changes, such as glass curtain walls and transparent structures, the depth estimation results are unstable due to failure to extract robust features. In this paper, we tackle this remaining challenge of the camera's difficulty in simultaneously reconstructing texture-poor glass curtain wall surfaces and texture-rich buildings.

The second challenge is the reconstruction granularity, which directly affects ray tracing accuracy. The 3D models used in contemporary multipath raytracing unanimously assume that each face of the building is rectangle, as exemplified by the widespread adoption of RANSAC-based planar segmentation proposed by Schnabel et al \cite{bib15}. However, most buildings in the real world and in particular dense urban environments like Lujiazui or Canary Wharf are bounded by polygons with more than or less than four corners. Inaccuracy in modeling the true boundaries of the building will in the end lead to inability to trace the rays correctly. Dai et al. adopted 3D-BoNet-based instance segmentation to decompose irregular building point clouds but struggled with over-segmentation of planar regions \cite{bib16}. Hackel et al. improved on the method of Schnabel et al. with contour-driven boundary detection \cite{bib17}, but failed to address adaptive parameterization of polygons with a variable number of vertices because their greedy contour simplification introduced jagged edges unsuitable for ray tracing. In this paper we will tackle this remaining problem by capturing globally optimized building polygonal shapes instead of simple and few regular large planes.

The last challenge is the balance between computational efficiency and accuracy. To lift off the hardware limitation, 3D Mapping Aided (3DMA) GNSS achieves quantitative modeling and suppression of multipath effects by fusing a high-precision 3D environment geometry model (covering the normal direction, curvature, and spatial topology of building surfaces) with the physical properties of signal propagation (including material dielectric constant, surface roughness, and polarization response). The shadow matching method utilizes building models to predict satellite visibility and achieves 3-meter accuracy in intersection scenarios \cite{bib18}, and subsequent research combines smartphone GNSS chips with machine learning classification to further extend dynamic scene adaptation \cite{bib19}. A typical ray-tracing technique proposed by Suzuki et al. simulates reflection paths and helps the particle filtering framework to effectively reduce the localization error of single-frequency receivers \cite{bib20} while the Advanced Ray Tracing (ART) algorithm reduces the computation time through a visual surface screening \cite{bib21}. Among the vision-assisted modeling methods, Skymask 3DMA uses a fisheye camera to segment the sky region to achieve NLOS culling, which improves the localization accuracy by up to 67 $\%$, but the edge reconstruction accuracy for near-ground building features is degraded due to their lens distortion effects,  directly reducing the reliability of GNSS multipath prediction \cite{bib22}. In summary, existing methods generally face the contradiction between modeling accuracy and real-time: high-precision ray tracing requires minute-level computation, while real-time algorithms suffer from accuracy loss due to model simplification \cite{bib21}. In addition, the problem of insufficient adaptability to complex building structures is prominent, and existing methods assume vertical wall surfaces or homogeneous materials, and lack effective modeling of curved glass curtain walls \cite{bib23} that can be viewed as an extended case of surface composed of polygons.

To bridge these gaps, this study proposes a framework that integrates real-time visual 3D reconstruction with optimized ray tracing for multipath estimation. Our approach employs a multi-plane fitting algorithm to reduce the complexity of the 3D point cloud maps, efficiently reconstructing and traversing an abundance of reflective surfaces in complex urban settings. By identifying the fact that multipath incurring on street vehicles is largely caused by reflections from wall sections between 10-60 meters height, we screen off the effects of dynamic obstacles (e.g., pedestrians and vehicles) while addressing the limitations of the camera's field of view in capturing high-rise buildings. Key innovations of this work include:

\begin{itemize}
\item To address the challenge of robust visual feature extraction, an orthogonal feature fusion framework is proposed to integrate SuperPoint and Scale-Invariant Feature Transform (SIFT) descriptors through feature space orthogonality analysis. The architecture implements a multi-hypothesis fusion mechanism tailored for complementary building materials, which adaptively enhances texture-poor semantic features and texture-rich geometric features by eliminating redundancy in orthogonal subspaces, leading to dynamic complementary enhancement.
\item To address the challenge of reconstruction granularity, we propose method to recover arbitrary shaped building boundary by refining point cloud segmentation. Unlike traditional rectangle- or triangle-based methods, our approach establishes a polygon-oriented surface reconstruction framework through parametric boundary regularization, which effectively preserves sharp and curved geometric features in complex architectural forms while maintaining topological consistency.
\item To tackle the temporal complexity of ray-tracing-based multipath mitigation, a complexity control strategy is proposed to achieve significant computational reductions through hierarchical feature filtering while maintaining the original performance metrics, ensuring an optimal balance between computational efficiency and reconstruction accuracy.
\item The proposed framework, composed of the above three proposed methods, is validated by extensive field tests in the Lujiazui area of Shanghai, where high-rise buildings and glass facades create a challenging multipath-inducing environment.
\end{itemize}

The rest of the paper is organized as follows: Section II describes the proposed framework. Section III details the mitigation pipeline, including the proposed orthogonal feature fusion method, arbitrary shaped reflecting surface reconstruction  and the temporal complexity control strategies for efficient multipath mitigation by tracing the rays bouncing off the reconstructed street building surface. Section IV presents the experimental results and analysis. Section V shows the results, with analysis and discussion.  Finally, Section VI includes a summary of the study and discusses future research directions.

\section{FRAMEWORK ARCHITECTURE}

\begin{figure*}
\centerline{\includegraphics[width=40pc]{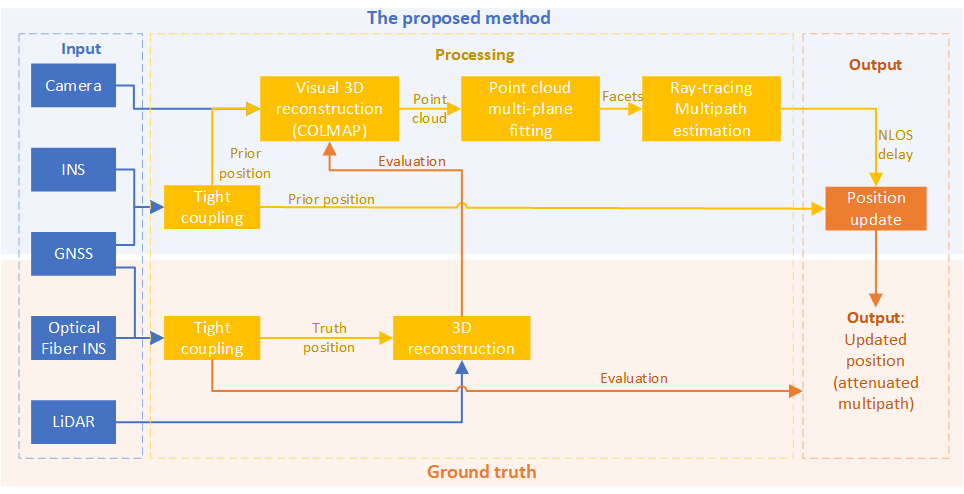}}
\caption{Framework architecture}
\label{overview}
\end{figure*}

The algorithm consists of two main phases, as outlined in Fig. \ref{overview}. In the first phase, we execute the COLMAP algorithm to obtain the initial mapping, and thus the fitted planar map, to avoid a large number of point clouds, decrease the plane fitting noise, increasing the multipath estimation speed. We do point cloud multi-plane fitting before multipath estimation. Then during the multipath estimation phase, we perform ray tracing to estimate the NLOS delay, and perform localization updates. A MEMS INS and GNSS are tightly coupled to provide a priori position for 3D reconstruction and multipath estimation. Meanwhile, a high-precision fiber-optic INS and GNSS are tightly coupled to provide ground truth for evaluating the accuracy of the multipath estimation algorithm. In addition, this GNSS+INS is combined with a LiDAR to evaluate the accuracy of the 3D reconstruction, as part of an ablation test or sensitivity test for the proposed method.

\subsection{Definition of Coordinate System}

\begin{figure}
\centerline{\includegraphics[width=18.5pc]{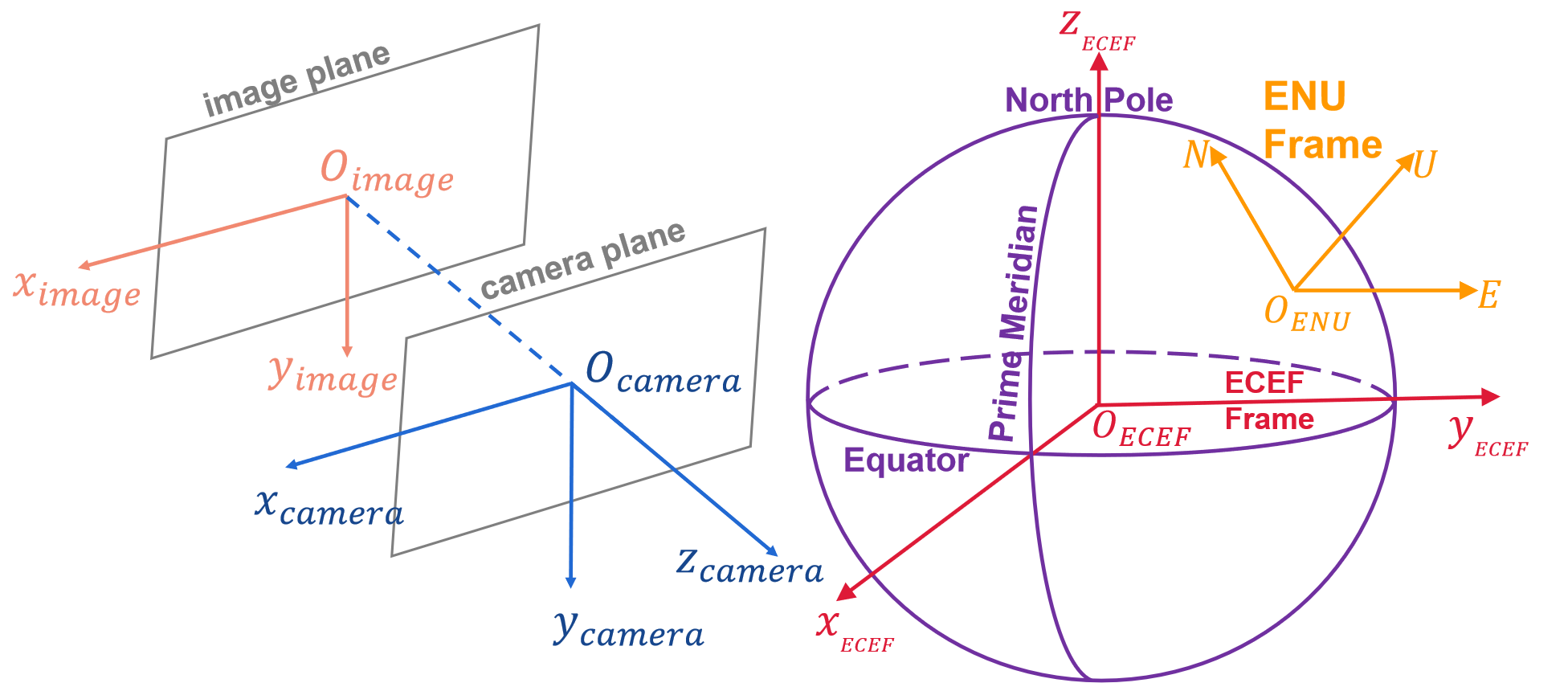}}
\caption{Coordinate system definition}
\label{frame}
\end{figure}

The coordinate system that will be covered in the paper is illustrated in Fig. \ref{frame}. For camera, we will utilize image coordinate system (i-frame) and camera coordinate system (c-frame) shown as the left panel of Fig.  \ref{frame}. For global positioning, ECEF coordinate system (e-frame) and ENU coordinate system (enu-frame) are defined in the right panel of Fig.  \ref{frame}.

\begin{figure}
\centerline{\includegraphics[width=16.5pc]{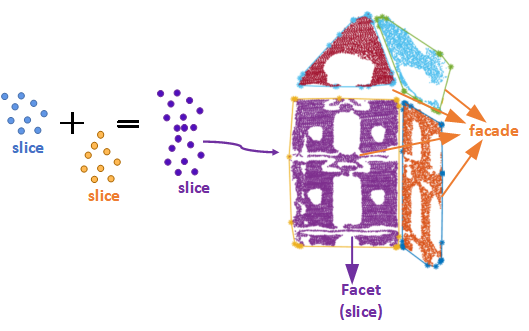}}
\caption{Nomenclature and examples}
\label{facet}
\end{figure}

\subsection{Nomenclature in Reconstruction}

For the current 3D reconstruction problem whose result is necessary for ray tracing and multipath estimation, a system of 2D geometric entities is defined here in Fig. \ref{facet}, including \textit{slice}, \textit{facet} and \textit{façade}, facilitating problem statement and the ensuing analyses. Slice is used in point segment, denotes the set of point cluster. Facet and façade are used in multipath estimation. Before delimiting these two words, we need to first define \textit{plane}. It denotes a surface without boundary. Facet denotes a surface with boundary. And facade denotes all the wall of a building; it includes several facets.

\section{THE METHODOLOGIES}

This section details the complete pipeline of 3D reconstruction, ray-tracing, and multipath estimation and correction. During the development, the proposed methods of orthogonal image feature fusion, fine-grained reconstruction through arbitrary shaped boundary recovery and finally the computational complexity control strategies are interspersed.
\subsection{COLMAP Round-Trip Reconstruction}
\begin{figure}
\centerline{\includegraphics[width=16.5pc]{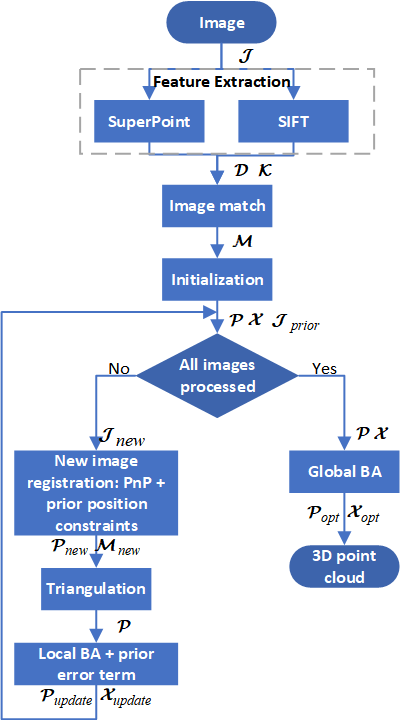}}
\caption{Incremental SfM flow chart}
\label{colmap}
\end{figure}

In this study, round-trip 3D reconstruction of the environment was performed based on the incremental Structure from Motion (SfM) method, COLMAP \cite{bib9}. Fig. \ref{colmap} shows the flow chart.

Firstly, for multi-material building structures in urban environments, we propose an orthogonal feature fusion framework for feature extraction in the scenes containing both texture-poor and texture-rich building surfaces to assist the COLMAP algorithm for 3D reconstruction.

\subsubsection{The Proposed Feature Extraction and Fusion Method}
This processing stage introduces a multi-material hypothesis framework, which is fundamentally different from the traditional single-material hypothesis. By coordinating SuperPoint and SIFT descriptors through orthogonal feature space constraints, we establish a cross-material adaptation mechanism. This architecture allows for simultaneous feature discrimination in texture-rich regions (e.g., brick walls) and texture-poor regions (e.g., glass curtain walls), ultimately enhancing the robustness of cross-material point cloud reconstruction.

The SuperPoint feature extraction first utilizes a VGG-style encoder for image size reduction to generate a feature map downscaled to 1/8 resolution of the original image (dimensionality of H/8×W/8×256). The feature map is decoded by parallel two-branch decoding: the feature point detection branch uses convolution and softmax activation to generate the feature point probability distribution map, and finally uses non-maximum suppression (NMS) to select the locally optimal keypoints; and the descriptor generation branch utilizes the same feature map through the convolution layer to output a 256-dimensional description vector, and performs sub-pixel refinement of the feature point locations by double and triple interpolation in the original image resolution. Then, the L2-normalization descriptors are used to obtain a uniform length description, and finally outputs a feature set containing precise coordinates and highly distinguishable descriptors \cite{bib8}.

It is the uniquely designed strategy combining pixel-based and descriptor-based method and also, the encoder that enables SuperPoint to achieve a strong characterization of texture-poor regions. Its generalized pattern of extracted contours learned from synthetic data in the training phase enables it to infer potential boundaries from semantic continuity of texture-poor surfaces (e.g., glass curtain walls), but feature extraction in texture-rich regions is inferior to conventional methods such as SIFT, as later exemplified in Fig. \ref{build}.

SIFT feature extraction first constructs a Gaussian pyramid for the input image, and detects the local extreme points as candidate keypoints by Difference of Gaussian (DoG) operator. Subsequently, the location and scale of the keypoints are determined by fitting a 3D quadratic function, while removing low-contrast keypoints and unstable edge response points. Next the keypoint principal directions are assigned based on the local gradient direction of the image. Finally, the descriptors are generated based on the gradient distribution in the rotationally normalized region, and the scale, direction, position, and description vector information are output after L2 normalization \cite{bib24}.

SIFT is based on a mathematical model of a Gaussian difference detector with a histogram of gradient directions, and its feature description is strongly sensitive to local pixel intensity mutations. This mechanism makes it perform more prominently in regions where rich texture is present (e.g., window frame edges), but inferior to learning-based methods such as SuperPoint in texture-poor regions, as later exemplified in Fig. \ref{build}.

Therefore, we propose to use a fused method: we extract SIFT features from texture-rich regions while SuperPoint features from texture-poor regions. We first define the captured images as $\mathcal{J}=\left\{\mathit{I}_1, \mathit{I}_2, \cdots, \mathit{I}_N \right\}$ where $\mathit{I_i}$ denotes the \textit{i}-th image, corresponding to the camera at \textit{i}-th epoch. The key points $\mathcal{K}=\left\{\mathbf{K}_1, \mathbf{K}_2, \cdots, \mathbf{K}_N \right\}$ is extracted where $\mathbf{K}_\mathit{i}=\left\{\mathbf{x}_\mathit{i}^1, \mathbf{x}_\mathit{i}^2, \cdots \right\}$  denotes the key points of the \textit{i}-th image. And descriptors are used to represent the characteristics of key points $\mathcal{D}=\left\{\mathbf{D}_1, \mathbf{D}_2, \cdots, \mathbf{D}_N \right\}$ where  $\mathbf{K}_\mathit{i} \in \mathbb{R}^{\mathit{m}\times\mathit{d}}$ denotes the descriptor matrix of the \textit{i}-th image (\textit{m} is the number of the features, \textit{d} is the descriptor dimension).

\subsubsection{Image Matching}
By using the descriptors $\mathcal{D}$ and the prior position $\mathcal{T}_{prior}=\left\{\mathbf{t}_{priot,1}, \mathbf{t}_{prior,2}, \cdots, \mathbf{t}_{prior,N} \right\}$  where $ \mathbf{t}_{prior,N}$ denotes the prior position of \textit{i}-th image in enu-frame, the overlapping part of two images is found and the key pair is matched. The prior position is obtained by the tightly coupled GNSS+MEMS INS. The output is a set of matching point pairs $\mathcal{M}=\{\mathbf{M}_{ij}| 1 \leqslant i < j < N \}$ where $\mathbf{M}_{ij}=\{(\mathbf{x}_i^k,\mathbf{x}_j^k) \} $ is the matched point pairs of images $\mathit{I_i}$ and image  $\mathit{I_j}$ in the image frame, \textit{i}-frame.

\subsubsection{Initialization}
The goal of the initialization phase is to build an initial 3D point cloud and camera pose using two-view geometry. In this stage, the input is matched point pair, camera intrinsics $\mathbf{K}$ and prior position information $\mathcal{T}_{prior}$. First, the optimal image pair is selected based on the number of matched points $\mathbf{M}_{ij}$ or prior position proximity, and the fundamental matrix $\mathbf{F}$ is estimated by using the matched point pair, and the internal parameter matrix is converted to the essential matrix

\begin{equation}
    \mathbf{E}=\mathbf{K}^\mathrm{T}\mathbf{F}\mathbf{K} .
\end{equation}
And the relative rotation matrix $\mathbf{R}$ and translation vector $\mathbf{t}$ of the two cameras are obtained by SVD decomposition. Triangulation is used to recover the initial 3D point $\mathbf{X}$ of matching feature point $\mathbf{x}$.

\subsubsection{Triangulation}
The goal of triangulation is recovering the 3D points $\mathcal{X}=\{\mathbf{X}_1,\mathbf{X}_2, \cdots \}$ in enu-frame.

For $i$-th image, the relationship among $k$-th 2D point $\mathbf{x}_i^k$, camera pose and corresponding 3D point $\mathbf{X}_j$ are

\begin{equation}
    \mathbf{x}_\mathit{i}^\mathit{k} \times \mathbf{K} \left[ \mathbf{R}_\mathit{i} | \mathbf{t}_\mathit{i} \right] \mathbf{X}_\mathit{j} = \mathbf{0} .
\end{equation} 

Then, after establishing homogeneous equations and SVD, 3D points can be recovered.
\subsubsection{PnP}
The goal of PnP is estimating the camera pose by $n$ pairs of 3D-2D point correspondences.

The initial three-dimensional point cloud is generated by linear triangulation
\begin{equation}
    \gamma_i \mathbf{x}_\mathit{i}^\mathit{k} = \mathbf{K} \left[ \mathbf{R}_\mathit{i} | \mathbf{t}_\mathit{i} \right] \mathbf{X}_\mathit{i}^k 
\end{equation}
where $\gamma_i$ denotes the scale and $\mathbf{X}_\mathit{i}^k$ denotes the 3D point in enu-frame corresponding to $\mathbf{x}_\mathit{i}^\mathit{k}$.
Then, Direct Linear Transform (DLT) method are applied by establishing homogeneous equations and SVD to solve $\mathbf{R}$ and $\mathbf{t}$ which are the pose of the camera.

\subsubsection{Prior position constraints}
Known camera positions can be added as constraints to the said SfM optimization problem. Assuming the prior position of the \textit{i}-th image is $\mathbf{t}_{prior,i}$, then the constraint term can be expressed as: 
\begin{equation}
    E_{prior}(\mathbf{t}_i)=\| \mathbf{t}_i-\mathbf{t}_{prior,i} \|^2
\end{equation}
where $\mathbf{t}_i$ is the optimization variable (camera position or translation). This constraint term is used to bracket the optimizable range of the camera position to avoid reconstruction failure due to excessive deviation of the initial value.

\subsubsection{Bundle Adjustment}
If the prior position is known, the cost function of the bundle adjustment will add a prior constraint term:
\begin{equation}
\begin{array}{c}
    \min_{\{\mathbf{R}_i, \mathbf{t}_i\}, \{\mathbf{X}_j\}}
    \left(
    \sum_{i,j} \rho (\| \pi (\mathbf{R}_i \mathbf{X}_j+\mathbf{t}_i )- \mathbf{x}_{ij} \|^2)\right.\\
    \left.+\alpha \sum_i E_{prior}(\mathbf{t}_i)  \right)
\end{array}
\end{equation}
where $\pi(\cdot)$  is the projection function, $\rho(\cdot)$ is the robust kernel function and  is the weight related to the GNSS/INS positioning accuracy.

\subsection{The Proposed Arbitrary-Shaped Boundary Recovery Method by Multi-Plane Fitting of Point Cloud}
\begin{figure}
\centerline{\includegraphics[width=16.5pc]{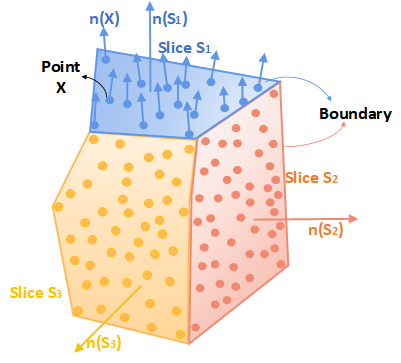}}
\caption{Schematic diagram of point cloud segmentation and normal estimation}
\label{slice}
\end{figure}

Point cloud segmentation can be formulated as a feature-based clustering task, as geometrically adjacent points on local surfaces typically exhibit similar normal orientations (Fig. \ref{slice}). We employ an enhanced P-Linkage clustering framework \cite{bib25} for point cloud segmentation, which differs from conventional 2D clustering in three aspects: (1) neighborhood construction via K-nearest neighbors (KNN), (2) surface flatness as the feature metric, and (3) normal vector deviation as the similarity criterion. The workflow is illustrated in Fig. \ref{point_chart}.
\subsubsection{Normal Estimation}
Based on the KNN strategy \cite{bib26}, the spatial topological relationship of the point cloud is constructed, and the local surface geometric features are calculated by principal component analysis (PCA).  

\begin{figure}
\centerline{\includegraphics[width=16.5pc]{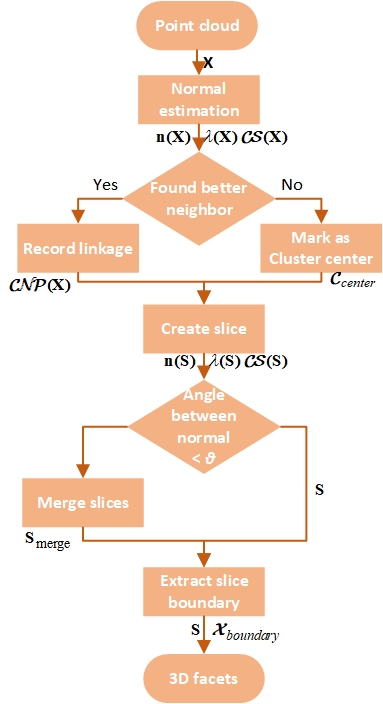}}
\caption{Multi-plane fitting of point cloud flow chart}
\label{point_chart}
\end{figure}

The specific implementation is divided into three stages: firstly, the k-d tree spatial index structure is established to screen the distance-ordered $K$ neighboring points for each sampling point. For each data point $\mathbf{X}_{i}$, its KNN is found and recorded as $\mathcal{X}_{KNN}$. Secondly, the covariance matrix $\Sigma$ is constructed by selecting the first $K$/2 nearest neighboring points as

\begin{equation}
\begin{array}{c}
    \Sigma = \frac{1}{K/2} \sum_{i=1}^{K/2}(\mathbf{X}_{\mathrm{KNN},i}-\overline{\mathbf{X}}_{\mathrm{KNN}})(\mathbf{X}_{\mathrm{KNN},i}-\overline{\mathbf{X}}_{\mathrm{KNN}})^T ,\\
    i \leqslant K/2 , \mathbf{X}_{\mathrm{KNN},i} \in \mathcal{X}_{\mathrm{KNN}}
\end{array}
\end{equation}
where $\overline{\mathbf{X}}_{\mathrm{KNN}}$  denotes the mean vector of the first $K$ / 2 data points in $\mathcal{X}_{\mathrm{KNN}}$ (the centroid of the selected neighbors). By solving the characteristic equation

\begin{equation}
    \lambda \mathbf{v} = \Sigma \mathbf{v}
\end{equation}
the eigenvector corresponding to the smallest eigenvalue is obtained and used as the normal vector $\mathbf{n}_i \in \left\{\mathbf{v} \right\}$ as shown in Fig. \ref{point_chart}, and its corresponding eigenvalue $\lambda_0$ is used to characterize the surface flatness.

To improve the noise immunity, the maximum consistency criterion is used to screen the valid interior points: the set of orthogonal distances $\left\{d_0^k \right\}_{k=1}^K$ from the $K$ neighboring points to the fitting plane is calculated as $\mathcal{N}_{OD}=\left\{d_0^k \right\}_{k=1}^K$, and the anomaly threshold was calculated by the median absolute deviation method:
\begin{equation}
    MAD = \mu \times median_{d_0^k \in \mathcal{N}_{OD}} |d_0^k-median(\mathcal{N}_{OD})|
\end{equation}
where $\mu = 1.4826$, which is defined by Lu et al \cite{bib25}.

The inliers, also known as the consistent set $\mathcal{CS}$ satisfying:

\begin{equation}
    R_z = \frac{|d_0^k-median(\mathcal{N}_{OD})|}{MAD}
\end{equation}
less than a threshold 2.5 \cite{bib27}. Finally, through all these filtering, we obtain the normal vector $\mathbf{n}(\mathbf{X}_i)$, flatness $\lambda(\mathbf{X}_i)$ and consistency set $\mathcal{CS}(\mathbf{X}_i)$ of each point. 

\subsubsection{Linkage Building}
This step is to construct point cloud topological connectivity relationships based on geometric features: for each sampling point  $\mathbf{X}_i$, search for the point with better flatness $\lambda(\mathbf{X}_i)$ and smallest normal vector deviation in its coherent neighborhood as the closest neighboring point (CNP) $\mathcal{CNP}(\mathbf{X}_i)$), which is recorded in a lookup table $\mathbb{T}$. The table contains fields such as point index, CNP identification, cluster number and eigenvalue. In order to ensure that the clustering center has a high degree of flatness and suppress the interference caused by surface changes or noise, 

\begin{equation}
    {th}_{\lambda}= \overline{\lambda}+\alpha_{\lambda}
\end{equation}
is adopted as a judgment threshold, where $\overline{\lambda}$ is the global flatness mean and $\alpha_{\lambda}$ is the standard deviation. Points with flatness below this threshold are selected as candidate clustering center $\mathbf{C}_{\mathrm{center}}$.

\subsubsection{Slice Creating}
To create the surface slices, the clusters $\mathcal{C}$ are firstly formed by searching the lookup table $\mathbb{T}$. For each cluster center $\mathcal{C}_{\mathrm{center}}$, collect the data points that are directly or indirectly connected with it. Set the scale threshold (e.g., 200 points) according to the actual scenario satisfying topological consistency while minimizing the number of slices to filter out outlier clusters. The retained clustered clusters are re-estimated with slice parameters. For each slice $\mathbf{S}_{i}$, we obtain its normal $\mathbf{n}(\mathbf{S}_{i})$, flatness $\lambda(\mathbf{S}_{i})$ and Consistent Set $\mathcal{CS}(\mathbf{S}_{i})$ in the same way as each data point.

\subsubsection{Slice Merging}
To realize complex surface reconstruction, a merging strategy based on normal vector constraints is proposed: if the neighboring slices $\mathbf{S}_{p}$ and $\mathbf{S}_{q}$ satisfy:
\begin{equation}
    \begin{array}{ccc}
   \ \ \ \ \ \ \ \ \ \exists \mathbf{X}_i \in \mathcal{CS}(\mathbf{S}_p) \ & and \ & \mathbf{X}_j \in \mathcal{CS}(\mathbf{S}_q), \\
    where \ \mathbf{X}_i \in \mathcal{CS}(\mathbf{X}_j) \ & and \ & \mathbf{X}_j \in \mathcal{CS}(\mathbf{X}_i)  
    \end{array}
\end{equation}
and the normal vector angle satiafies:
\begin{equation}
    arccos |\mathbf{n}(\mathbf{S}_p)^T \cdot \mathbf{n}(\mathbf{S}_q) | < \theta  
\end{equation}
then the slices are merged, where the threshold  is determined based on further analysis of planar tilting error, deferred to Section III.E. The parameter controls the surface continuity of the sensitivity.

\subsubsection{Slice Boundary Extraction}
For each segmented point cloud cluster, the associated planar surface is parameterized by a general plane equation $\beta x +\varepsilon y+\epsilon z +\tau=0$. Subsequently, all points within the cluster are orthogonally projected onto the corresponding fitted plane. To facilitate boundary extraction, the 3D coordinates of the projected points are transformed into a 2D local coordinate system defined on the planar surface. Following this dimensionality reduction, the convex hull vertices of the projected 2D point set are computed using Graham's scan algorithm \cite{bib28}. The method is as follows: 

Firstly, the point with the minimum \textit{y}-coordinate is selected as the pivot $\mathbf{X}_0$. If multiple points share the same minimum \textit{y}-coordinate, the point with the smallest \textit{x}-coordinate is chosen. This ensures $\mathbf{X}_0$ lies on the convex hull boundary. 

Secondly, all remaining points are sorted by their polar angles relative to $\mathbf{X}_0$ in counterclockwise order. Points with identical polar angles are further sorted by their Euclidean distances from $\mathbf{X}_0$, prioritizing closer points to avoid redundancy. The first two points in this sorted sequence, denoted $\mathbf{X}_1$ and $\mathbf{X}_2$, are pushed onto the stack along with $\mathbf{X}_0$. 

Finally, for each subsequent point $\mathbf{X}_j$ in the sorted list, let $\mathbf{E}$, $\mathbf{F}$, and $\mathbf{G}$ denote the second-top, top element of the stack, and the current point $\mathbf{X}_j$, respectively. Compute the cross product of vectors $\vec{EF}$ and $\vec{EG}$: 
\begin{equation}
    cross = (F_x - E_x)(G_y-E_y)-(F_y-E_y)(G_x-E_x) . 
\end{equation}

If $cross< 0$, $\mathbf{G}$ lies to the right of $\vec{EF}$. Pop $\mathbf{F}$ from the stack and re-evaluate the new top elements until the cross product becomes non-negative. If  $cross> 0$, $\mathbf{F}$ lies to the left of $\vec{EF}$. Push $\mathbf{G}$ onto the stack. If $cross=0$, $\mathbf{G}$ is collinear with $\mathbf{E}$ and $\mathbf{F}$, retain the farthest point from $\mathbf{X}_0$ to minimize redundancy. This process is iterated until all points have been processed, resulting in a stack containing the convex hull vertices in counterclockwise order. Points in the stack are the boundary of the slice, we denote bounded slice by facet. Record boundary points of each slice into set $\mathcal{X}_{boundary}$.

In this study, we performed multi-plane fitting on large-scale point clouds. For adjacent point clouds, the normal vector threshold $\theta$ was determined based on the allowed tilt error calculated in Section III.E. And a minimum allowable cluster size of 200 points was established to eliminate noise interference during point cloud merging. Each clustered point set subsequently underwent individual plane fitting. Polygon boundaries were then extracted from each fitted plane. This methodology achieved significant plane quantity reduction while improving real-time performance in multipath estimation.

\begin{figure*}
    \centering
    \begin{subfigure}{0.45\textwidth}
        \includegraphics[width=\textwidth]{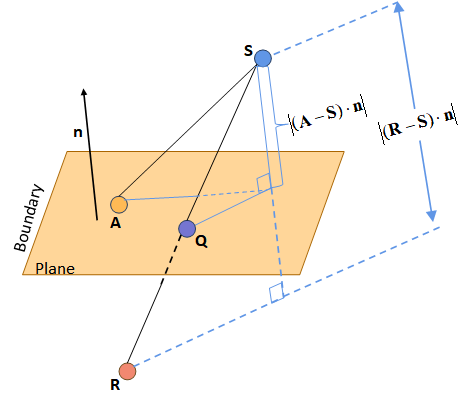}
        \caption{Schematic diagram for determining the intersection of propagation paths and blocking planes}
        \label{LOS_NLOS_a}
    \end{subfigure}
    \hfill
    \begin{subfigure}{0.45\textwidth}
        \includegraphics[width=\textwidth]{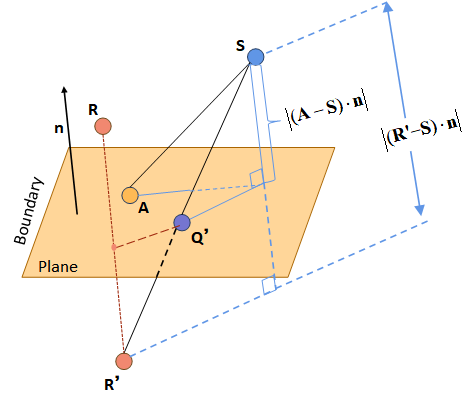}
        \caption{Schematic diagram for calculating the reflection point, mirror point, and reflection path}
        \label{LOS_NLOS_b}
    \end{subfigure}
    \caption{Schematic diagram for ray tracing}
    \label{LOS_NLOS}
\end{figure*}

\subsection{Ray Tracing}
The propagation delay of the satellite navigation signal in the multipath-rich area at the GPS L1 frequency is calculated using the method of three-dimensional ray tracing. Based on the geometric relationship along the path between the receiver and building facade, it is determined whether the propagation path is simply blocked or undergoes reflection. And it is with this information the reception type of the signal from the satellite is judged and the propagation delay is calculated and used for correction \cite{bib29}.

\subsubsection{Determining Signal Paths and Building Intersections}
In this paper, we determine whether the satellite signal will be blocked by a building by judging if a hypothetical ray mimicking radio frequency signal emitted from the satellite heading for the receiver intersects with the facade of the building. Fig. \ref{LOS_NLOS_a} illustrates the thinking. The figure represents the satellite for which the reflection path is calculated as $\mathbf{S}$; $\mathbf{A}$  represents an arbitrary point on the surface of the suspected facade, and $\mathbf{R}$ represents the receiver whose line-of-sight (LOS) path to the satellite $\mathbf{S}$ may be blocked. The inner product of the vector from the satellite to the point $\mathbf{A}$ on the plane and the normal vector $\mathbf{n}$ of the plane is the normal distance from the satellite to the plane. At the same time, the inner product of the vector from the satellite to the receiver and the normal vector of the plane is the distance between the two along the normal vector of the corresponding plane. The product of the ratio of the above two components and the vector from the satellite $\mathbf{S}$ to the receiver $\mathbf{R}$ is the point of intersection of the line segment of the satellite and the receiver, with the infinite plane formed by the extension of that plane and the hypothetical intersection point $\mathbf{Q}$, can be computed as
\begin{equation}
    \mathbf{Q}=\frac{\|(\mathbf{A}-\mathbf{S})\cdot \mathbf{n}\|}{\|(\mathbf{R}-\mathbf{S})\cdot \mathbf{n}\|}  \cdot (\mathbf{R}-\mathbf{S})+\mathbf{S} .
\end{equation}

If $\mathbf{Q}$ belongs to the facet, it means that the path from the satellite to the receiver has an intersecting relationship with one of the facets, and the transmitted signal is blocked by the building façade containing the facet, so that the type of the received signal is NLOS or blocked; if the intersection point doesn't belong to the facet, the including facade won't block the signal. If all planes do not obscure the signal transmission, the signal type at the receiving end is LOS+NLOS or LOS only, i.e., a direct path exists. 

To determine whether a point is inside a polygon (facet), we use the Angle Sum Method: connect the point $\mathbf{Q}$ with the vertices of the convex polygon in counterclockwise order and then calculate the directed angle formed by each pair of neighboring vertices to $\mathbf{Q}$. If all angles are in the same direction (e.g., counterclockwise) and summed up to 360°, then $\mathbf{Q}$ is inside the facet. If positive and negative angles cancel each other out and are summed up to 0°, then $\mathbf{Q}$ is outside the facet.
The algorithm needs to traverse all candidate facets and the complexity of the algorithm is $O(m)$, if the number of facets is set to $m$.

\subsubsection{Reflected Path Calculation}
A signal can also be reflected instead of blocked; thus it is necessary to calculate the hypothetical reflected path. In this regard, the method of calculating the three-dimensional mirror points and judging the intersection relationship of the plane is used to judge whether there exists a signal transmission path from the satellite $\mathbf{S}$ to the receiver $\mathbf{R}$ through the reflection of the plane. Fig. \ref{LOS_NLOS_b} shows a schematic diagram of the algorithm for calculating the reflection point and the mirror image point $\mathbf{R}'$, and judging whether there is a reflection path, where $\mathbf{n}$ is the unit normal vector of the plane. Traversing all the facets in the urban environment model, the receiver's mirror point $\mathbf{R}'$, the vector between the mirror point and the receiver, is the normal component of the vector from twice the receiver $\mathbf{R}$ to a point $\mathbf{A}$ on the plane. The normal component can be obtained from the inner product of the receiver to one point on the plane vector, and the normal vector. 

\begin{equation}
\left\{
\begin{array}{c}
(\mathbf{R}'-\mathbf{R}) \times \mathbf{n} = 0\\
\|\mathbf{R}'-\mathbf{A}\|=\|\mathbf{R}-\mathbf{A}\|
\end{array}
\right.
\end{equation}
or
\begin{equation}
\mathbf{R}'=\mathbf{R}+2\cdot ((\mathbf{A}-\mathbf{R})\cdot \mathbf{n})
\end{equation}
The intersection point $\mathbf{Q}'$ of the virtual transmitted signal from the satellite to the mirror point $\mathbf{R}'$ of the receiver and the extended plane, that is, the reflection point of the signal in this plane can be calculated as

\begin{equation}
    \mathbf{Q}'=\frac{\|(\mathbf{A}-\mathbf{S})\cdot \mathbf{n}\|}{\|(\mathbf{R}'-\mathbf{S})\cdot \mathbf{n}\|}  \cdot (\mathbf{R}'-\mathbf{S})+\mathbf{S} .
\end{equation}
If the intersection point $\mathbf{Q}'$ belongs to the facet, it indicates that there is a reflection transmission path reflected through the plane, and the reflection point is $\mathbf{Q}'$. If not, there is no reflection path induced by this plane.

\subsubsection{Reflection Delay Calculation}
If there is a reflection path, the reflection delay $d$ is the sum of the distance from the satellite to the reflection point and the distance from the reflection point to the receiver, minus the LOS distance from the satellite to the receiver, namely:
\begin{equation}
    d=\|\mathbf{Q}'-\mathbf{S}\|+\|\mathbf{R}-\mathbf{Q}'\|-\|\mathbf{R}-\mathbf{S}\| .
\end{equation}

\subsection{The Proposed Complexity Control Strategy 1: Reflection Point Height Constraint}
On the one hand, because buildings on both sides of an urban environment are obscured by objects such as trees and street signs, it is thus useless to reconstruct facades below certain heights. Also, signals reflected above or below certain heights cannot reach a receiver mounted on a vehicle running within the car lanes, most of which are several or tens of meters away from the buildings along the streets. Therefore, we would like to calculate the height range of the reflection points of the NLOS to determine the subset of the suspected reflection facets that will ultimately be responsible for multipath effects.

On the other hand, since the ray tracing method needs to traverse all the building planes, we would like to reduce the number of planes used for multipath estimation by estimating the concentrated heights of the computed reflection points, thus improving the estimation efficiency.
\begin{figure}
\centerline{\includegraphics[width=18.5pc]{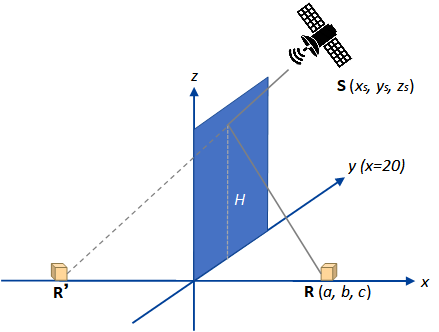}}
\caption{Reflection point, satellite and receiver geometry diagram}
\label{height error}
\end{figure}
Assume that the street width is 40 m which is the widest avenue in central Shanghai, implicating the smallest possible height constraint to be computed and vehicles are moving along the centerline of the street. Then let the position of the vehicle be $\mathbf{R}=(a,b,c)$, the position of the satellite be $\mathbf{S}=(x_s,y_s,z_s)$, and the position of the wall be $x=\pm20$. The hypothetical position of the vehicle due to reflection, i.e., the mirror point is $\mathbf{R}'=(\pm 40-a, b, c)$. Fig. \ref{height error}. shows the geometry when $x = 20$. The parametric equation from the mirror point to the satellite can be obtained:
\begin{equation}
\left\{
\begin{array}{ccc}
x&=&x_s+t(\pm 40-a-x_s)\\
y&=&y_s+t(b-y_s)\\
z&=&z_s+t(c-z_s)
\end{array}
\right.  .
\label{19}
\end{equation}

Since the movement of vehicle is assumed along the centerline, then we can estimate the height of the reflection point on the wall as 
\begin{equation}
H=\frac{20-x_s}{\pm40-a-x_s}(c-z_s)+z_s .
\label{20}
\end{equation}

\subsection{The Proposed Complexity Control Strategy 2: Maximum Allowable 3D Reconstruction Error Margin}
3D reconstruction has inevitable errors. For the purpose of estimating pseudorange multipath delay, it is unnecessary to reach AR/VR accuracy (down to centimeters), but where the line should be drawn is an unsolved problem. The concept of tolerable reconstruction error is proposed, and we propose method to compute it. Therefore, we consider analyzing the translation and tilt errors allowed for 3D reconstruction that can still accurately help us judge NLOS.

To address the spatial distributive characteristics of point clouds existing in 3D sparse reconstruction and their influence on the geometric modeling accuracy, we propose a reflection path discrimination mechanism based on fault-tolerant threshold. Due to the neighbor spacing property caused by sparse point cloud sampling, the geometric accuracy of the facet boundary is constrained by the maximum spacing in the null space between neighboring point clouds, and this phenomenon will directly affect the robustness of the discrimination between LOS and NLOS signal propagation paths. For this reason, we introduce the key parameter $l$ as the airspace tolerance threshold (see IV.B for details of the specific parameter optimization method), and establish the tolerance deviation range with the real reflection point as the center and $l$ as the effective radius. The model is mathematically equivalent to allowing the reflection plane to have a translation error of $e_{max}$ in the normal vector direction and a rotational deviation of $\theta$ in the tangent plane direction.

\begin{figure*}
    \centering
    \begin{subfigure}{0.45\textwidth}
        \includegraphics[width=\textwidth]{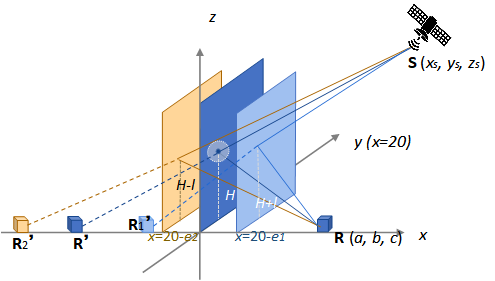}
        \caption{Geometry of wall translation error}
        \label{error_range_a}
    \end{subfigure}
    \hfill
    \begin{subfigure}{0.45\textwidth}
        \includegraphics[width=\textwidth]{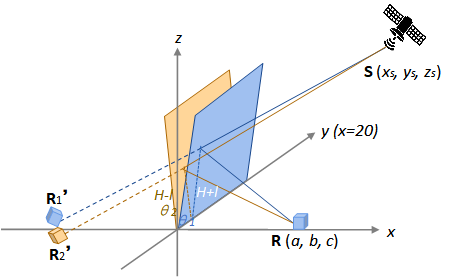}
        \caption{Geometry of wall tilt error}
        \label{error_range_b}
    \end{subfigure}
    \caption{Geometry of wall error}
    \label{error_range}
\end{figure*}

\subsubsection{3D Reconstruction Translation Error Margin}
Assume that the position of the receiver is $\mathbf{R}=(a,b,c)$; the real reflection point height is $H$ and the allowable reflection point height limit is $[H-l,H+l]$.  Define $e_1$ as the translation error when the reflection point height is $H+l$ and define $e_2$ as the translation error when the reflection point height is $H-l$. The geometry is shown in Fig. \ref{error_range_a}.

When the position of the received signal is $(a,b,c)$, assuming the wall position $x = 20$, the height of the reflection point is obtained according to Equation (20). The positions of the mirror points are $\mathbf{R}_1'=(40+2e_1-a,b,c)$ and $\mathbf{R}_2'=(40-2e_2-a,b,c)$, respectively. And according to (\ref{19}), we can get following equation:
\begin{equation}
\left\{
\begin{array}{ccc}
\frac{H+l-z_s}{c-z_s}&=&\frac{20+e_1-x_s}{2(20+e_1)-a-x_s}\\
\frac{H-l-z_s}{c-z_s}&=&\frac{20+e_2-x_s}{2(20+e_2)-a-x_s}
\end{array}
\right.  .
\end{equation}  
The maximum translation error value allowed for the 3D reconstructed wall planes can be obtained from the calculations:
\begin{equation}
\begin{array}{ccc}
e_{max}&=&e_1+e_2\\
&=&\frac{(H+l)(x_s+a)-cx_s-z_sa}{2H+2l-c-z_s}\\
&+&\frac{(H-l)(x_s+a)-cx_s-z_sa}{2H-2l-c-z_s}
\end{array}
.
\label{22}
\end{equation}

\subsubsection{3D Reconstruction Tilt Error Margin}
Once again, assume that the position of the antenna phase center is $\mathbf{R} =(a,b,c)$, the real reflection point height is $H$ and the allowable reflection point height limit is $[H-l, H+l]$. The geometry is shown in Fig. \ref{error_range_b}.

Assuming the wall position $x = 20$, the height of the reflection point can be obtained according to (\ref{20}). If the angle of inclination of the wall is $\theta$, the height of the reflection point is $(H\pm l)cos\theta$. The positions of the mirror points are $\mathbf{R}_1'=(20-(a-20)cos^2\theta_1 , b,c+2(a-20)sin^2\theta_1)$ and $\mathbf{R}_2'=(20-(a-20)cos^2\theta_2 , b, c-2(a-20)sin^2\theta_2)$, respectively.

Then the maximum tilt error allowed for 3D reconstruction of the wall plane can be calculated as:
\begin{equation}
\begin{array}{ccc}
\theta & = &\theta_1+\theta_2\\
& \approx  & \frac{(H+l)(40-x_s-a)+(x_s-20)c-z_s(20-a)}{(c-z_s)(H+l)}\\
& +  & \frac{(H-l)(40-x_s-a)+(x_s-20)c-z_s(20-a)}{(c-z_s)(l-H)}
\end{array}  .
\label{23}
\end{equation}

\section{Experiment}
\begin{figure}
\centerline{\includegraphics[width=18.5pc]{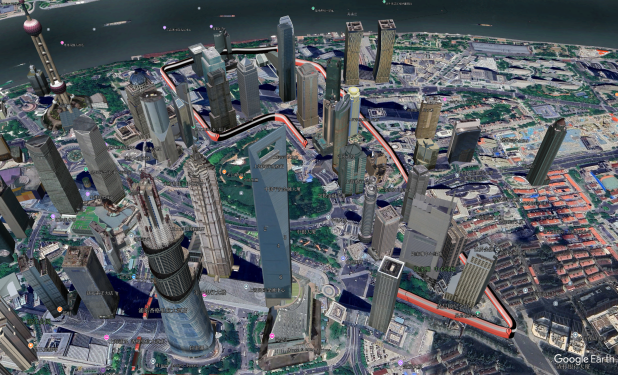}}
\caption{Experimental scenarios and experimental routes}
\label{scena}
\end{figure}

The experiments were conducted in Lujiazui, Shanghai, with a route traversing an area densely populated with high-rise buildings with reflective glass surfaces, shown as Fig. \ref{scena}.

\begin{figure}
\centerline{\includegraphics[width=18.5pc]{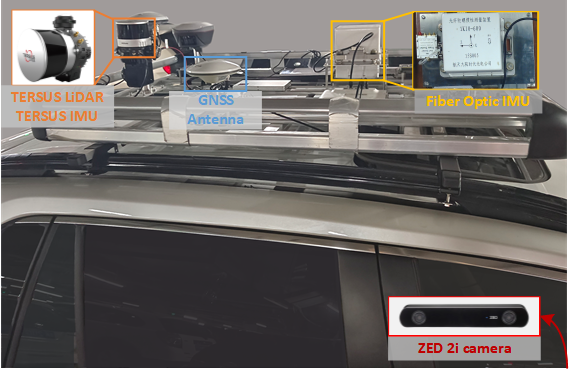}}
\caption{Experimental equipment}
\label{exper}
\end{figure}

Our experiments were conducted with TERSUS David 30 GNSS receiver, TERSUS inertial guidance, TERSUS LiDAR, ZED 2i camera, fiber optic inertial guidance. The important parameters or settings of these sensors are shown in Table \ref{tab1}. These devices were fixed on top of a Toyota RAV4 as shown in Fig. \ref{exper} and were pre-calibrated before the experiment using open-source tools \cite{bib30}.

\begin{table}
\caption{Equipment parameter}
\label{tab1}
\tablefont
\begin{tabular*}{20pc}{@{}p{4pc}p{6pc}p{7pc}@{}}
\hline
\textbf{Equipment} & \textbf{Function} & \textbf{Parameter} \\
\hline
GNSS Antenna & Positioning for both measurement and truth value &
Pseudorange observation accuracy: 10\,cm \newline
Carrier phase observation accuracy: 1\,mm \\
Tersus IMU & Measurement: Prior position &
Bias stability: $0.3^{\circ}/hr$ \newline
Noise density: $0.15^{\circ}/\sqrt{h}$ \\
ZED 2i Camera & Measurement: 3D reconstruction &
Focal length: 2.1\,mm \newline
Field of view: $110^{\circ}(H)\times70^{\circ}(V)\times120^{\circ}(D)$ \\
Fiber Optic IMU & Positioning truth value &
Bias stability: $0.1^{\circ}/hr$ \newline
Noise density: $0.05^{\circ}/\sqrt{h}$ \\
Tersus LiDAR & Mapping truth value &
Range: 120\,m \newline
Angular resolution: $0.18^{\circ}(H)\times1^{\circ}(V)$ \\
\hline
\end{tabular*}
\end{table}

\subsection{Reflection Point Height Simulation}
\begin{figure}
\centerline{\includegraphics[width=18.5pc]{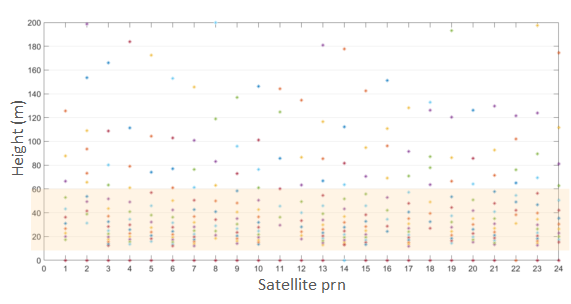}}
\caption{Height distribution of reflection points on the wall during one cycle of the simulated satellite}
\label{NLOS height}
\end{figure}

First, we evaluated the height constraint at a typical position $(31.24416^{\circ}N,\ 121.50347^{\circ}E)$, which generates multipath, by simulation according to Section III.D. We simulated traversing all the satellites in one orbital period and set the elevation angle $>\ 30^{\circ}$ . And we found that 63$\%$ of the reflection points are concentrated in the range of 10 m - 60 m above ground level, shown as Fig. \ref{NLOS height}. This range, therefore, provides the location where we constrain 3D reconstruction and multipath estimation.

\subsection{3D Reconstruction Point Cloud Distance Analysis}
\begin{figure}
\centerline{\includegraphics[width=18.5pc]{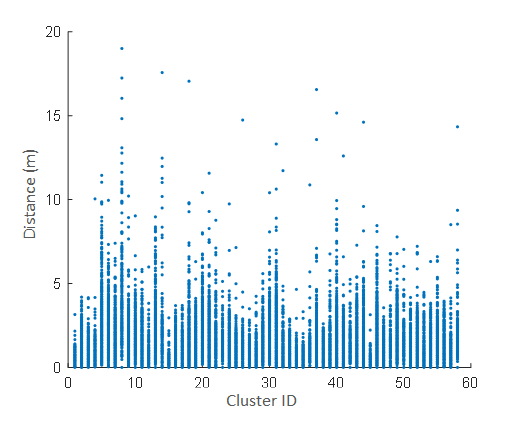}}
\caption{Distance statistics between neighboring point clouds in each cluster}
\label{distance statistic}
\end{figure}

\begin{figure}
\centerline{\includegraphics[width=18.5pc]{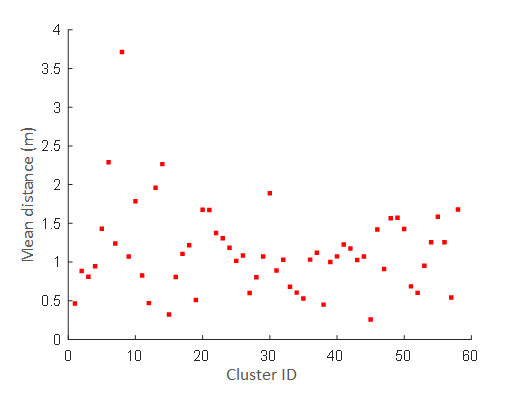}}
\caption{Mean distance statistics between neighboring point clouds in each cluster}
\label{cluster distance}
\end{figure}

In this part, we computed the distances between neighboring points of the 3D sparse reconstruction. After the point extraction, the point cloud is segmented to different clusters.
Fig. \ref{distance statistic} and Fig. \ref{cluster distance} count the distance between neighboring point clouds in each cluster and the mean distance between neighboring point clouds in each cluster. Table \ref{tab2} shows the statistics of mean and median of the point cloud distance. It includes the mean and median of all point cloud distance and the mean and the median of the mean of each cluster’s point cloud distance.

\begin{table}[htbp]
\centering
\caption{Point cloud distance statistics}
\label{tab2}
\begin{tabular}{|l|c|}
\hline
\textbf{Statistics} & \textbf{Value (m)} \\
\hline
Distance mean of all points & 0.9716 \\
Distance median of all points & 0.7691 \\
Mean of all clusters’ distance mean & 1.1448 \\
Median of all clusters’ distance mean & 1.0723 \\
\hline
\end{tabular}
\end{table}

Then, we will choose one of the statistics as threshold for section III.E, to evaluate the allowable reconstruction error. Since the reconstruction area encompasses both texture-rich and texture-poor building surfaces, there exists a density variance in the reconstructed point clouds. Due to the dense point cloud distribution in the texture-rich regions and the sparse characteristics in the texture-poor regions, directly adopting the distance statistics of the entire point cloud would lead to the results being dominated by the high-density areas. Therefore, by employing a phased statistical strategy, where the local mean of each point cloud cluster is calculated first and then the global statistical analysis is carried out, the comprehensiveness of the evaluation results can be better ensured. As shown in Fig. \ref{cluster distance}, significant outliers ($>$3.5 m) were discovered, and such outliers have obvious interferences on the arithmetic mean statistics. Comprehensive analysis indicates that the median statistics based on the mean of clusters can evenly reflect the distance distribution characteristics of point clouds in different density regions. Hence, it is established as the threshold  for determining the allowable error range in section IV.C and IV.D.

\subsection{Translation Error Margin}
To obtain the numerical value of the allowable reconstruction error margin (section III. E(1)), we evaluated at position $(31.24416^{\circ}N,\ 121.50347^{\circ}E)$ the range of reflection plane translation errors allowable for 3D reconstruction, according to (\ref{22}). We simulated traversing all the satellites in one orbital period and found the allowable reflection plane error. 

\begin{figure}
\centerline{\includegraphics[width=18.5pc]{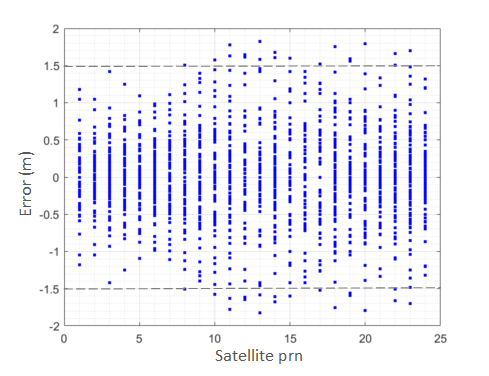}}
\caption{ Translation error distribution of wall during one orbital period of the simulated satellite}
\label{tran_error}
\end{figure}

Fig. \ref{tran_error} shows the distribution of the allowable translation error when the height limit is $[H-l,H+l]$. We find more than 99 $\%$ error $l$ is concentrated in the range of -1.5 m to 1.5 m. Therefore, the allowable translation error is set as 3 m.

\subsection{Tilt Error Margin}
\begin{figure}
\centerline{\includegraphics[width=18.5pc]{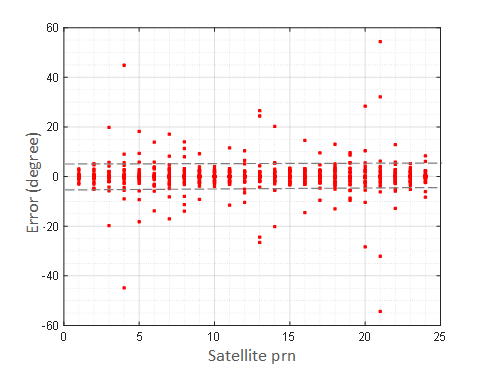}}
\caption{Tile error distribution of wall during one orbital revisit period of the simulated satellite}
\label{tilt_error}
\end{figure}

To obtain the numerical value of the allowable reconstruction error margin (section III. E(2)), we evaluated at position $(31.24416^{\circ}N,\ 121.50347^{\circ}E)$  the range of reflection plane tilt errors allowable for 3D reconstruction, according to (\ref{23}). We simulated traversing all the satellites in one orbital revisit period and found the reflection plane error. Fig. \ref{tilt_error} shows the distribution of the allowable translation error when the height limit is $[H-l,H+l]$. We find more than 99 $\%$ error $l$ is concentrated in the range of $-5^{\circ}$ to $5^{\circ}$. Therefore, the allowable rotational error margin is set as $10^{\circ}$.

\section{Results and Discussion}
\subsection{Results}
\begin{figure*}
    \centering
    \begin{subfigure}{0.45\textwidth}
        \includegraphics[width=\textwidth]{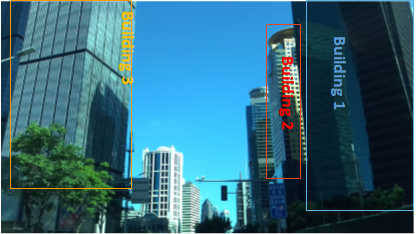}
        \caption{}
        \label{build_a}
    \end{subfigure}
    \hfill
    \begin{subfigure}{0.25\textwidth}
        \includegraphics[width=\textwidth]{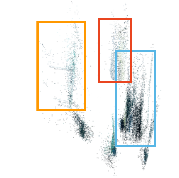}
        \caption{}
        \label{build_b}
    \end{subfigure}
    \hfill
    \begin{subfigure}{0.25\textwidth}
        \includegraphics[width=\textwidth]{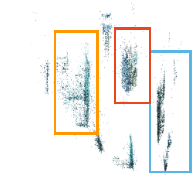}
        \caption{}
        \label{build_c}
    \end{subfigure}
    \caption{Comparative analysis of 3D reconstruction ((a) Experimental scene, (b) Reconstruction result based on SuperPoint feature extraction, (c) Reconstruction result base on SIFT feature extraction.)}
    \label{build}
\end{figure*}

We selected a typical road section dominated by glass curtain wall buildings as the experimental scene (shown in Fig. \ref{build_a}), and collected an image dataset. First, we used this dataset for 3D reconstruction comparative analysis between SuperPoint and SIFT feature extraction methods, respectively. The experimental scene contains three buildings with two significant feature differences: Building 1 (labeled by the blue box) is a texture-poor glass curtain wall structure, while Building 2 (red box) and Building 3 (yellow box) are texture-rich traditional building structures.

Fig. \ref{build_b} and \ref{build_c} compare 3D reconstruction results between SuperPoint and SIFT, respectively. The experimental results show that the feature extraction method has a significant impact on the reconstruction quality. For the texture-poor glass curtain wall structure (Building 1), the reconstruction results based on the SuperPoint feature extraction method are more fine-grained. The SuperPoint method shows better geometric detail restoration ability, and its reconstructed model exhibits higher completeness in the detailed features such as curtain wall joints and façade transitions; while for the texture-rich Building 2 and Building 3, the reconstruction results based on the SIFT feature extraction method are more fine-grained, especially in the masonry texture and other detail parts of the reconstructed point cloud is more intensive.

\begin{figure*}
    \centering
    \begin{subfigure}{0.4\textwidth}
        \includegraphics[width=\textwidth]{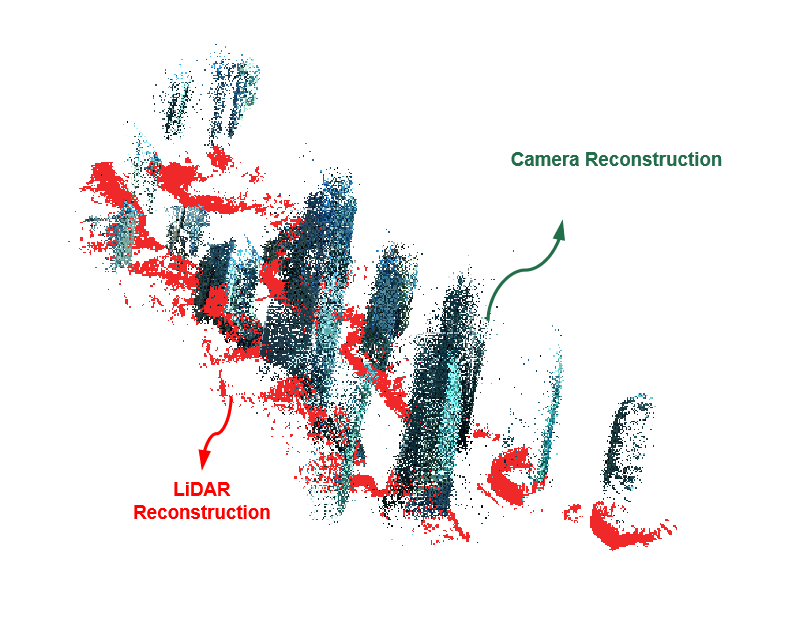}
        \caption{}
        \label{point_result_a}
    \end{subfigure}
    \hfill
    \begin{subfigure}{0.28\textwidth}
        \includegraphics[width=\textwidth]{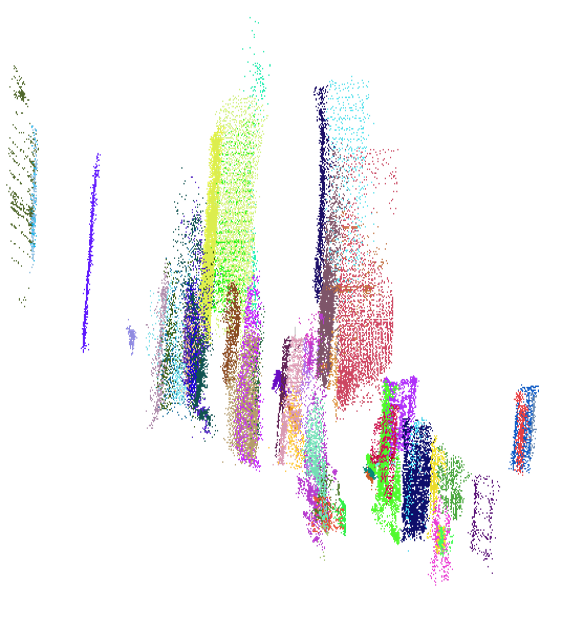}
        \caption{}
        \label{point_result_b}
    \end{subfigure}
    \hfill
    \begin{subfigure}{0.28\textwidth}
        \includegraphics[width=\textwidth]{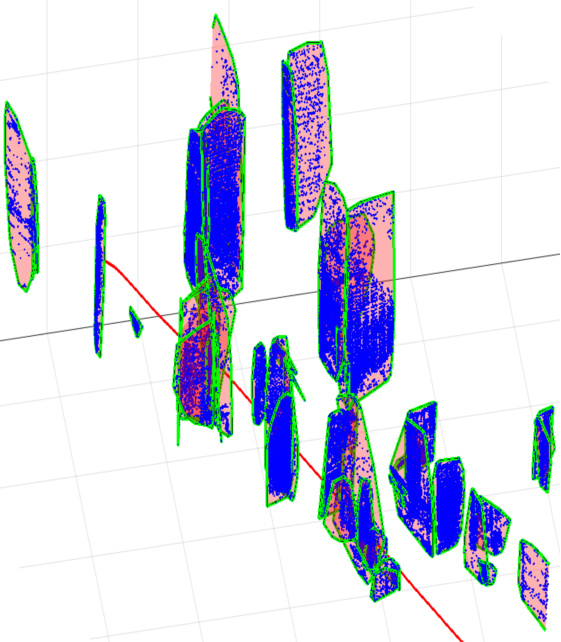}
        \caption{}
        \label{point_result_c}
    \end{subfigure}
    \caption{Reconstruction result ((a)  Camera (green points) and LiDAR (red points) reconstruction result map, (b) Point cloud segmentation, (c) 3D planar map based on point cloud multiplanar fitting with experimental path (red line).)}
    \label{point_result}
\end{figure*}

\begin{figure}
\centerline{\includegraphics[width=18.5pc]{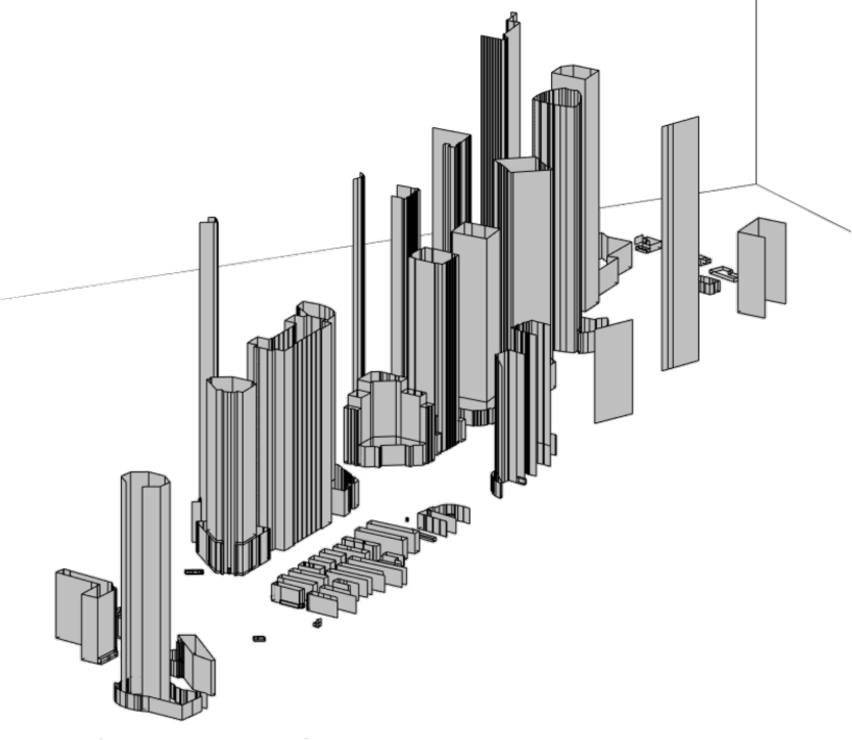}}
\caption{Pre-constructed offline simulation map}
\label{simulation}
\end{figure}

Fig. \ref{point_result_a} shows the 3D reconstruction results of all the buildings along a 4-minute ride in Lujiazui and the comparison with the ground truth obtained using LiDAR reconstruction. According to the point-to-plane error analysis which calculated distance statistics measured between the LiDAR point to camera-only reconstructed surface, the average error of 3D reconstruction is 2.4 m. The main reason for this error is the limitation of the LiDAR field of view (although the LiDAR sensor is already a state-of-the-art one), which reconstructs mainly the lower and close-to-vehicle area, while the higher part and the lower section have structural changes and are not a complete plane.  Therefore, using only the plane of the lower half of the building as the truth value for comparison is bound to result in errors. However, this mean error meets the allowable 3D reconstruction error margin analyzed in III.E (evaluated as a 3m upper bound in Section IV.C), and therefore, this modeling result is basically sufficient for modeling needs for multipath estimation. And Fig. \ref{simulation} shows the existing 3D building models extracted from google map (for brevity, we call it 'simulation map') of the experimental area. Comparing it with Fig. \ref{point_result_a}, we can find that our proposed methods satisfy the reconstruction coverage of the main buildings significantly important in calculating signal reflection or blockage.  And the simulation map also will be used in multipath estimation to compare with the proposed method.

Fig. \ref{point_result_b} shows the results of point cloud segmentation clustering based on the reconstructed point cloud map in Fig. \ref{point_result_a}, where point clouds within the same cluster are marked with the same color, and point clouds in different clusters are distinguished by color differences.

Fig. \ref{point_result_c} shows the 3D planar map of the point cloud after multi-plane fitting (section III.B), where the point clouds within the same cluster are fitted to the same plane. The point cloud within the cluster is projected onto the corresponding plane and the plane boundary is formed by convex polygons. The red line in the figure indicates the vehicle trajectory.

\begin{figure}
\centerline{\includegraphics[width=18.5pc]{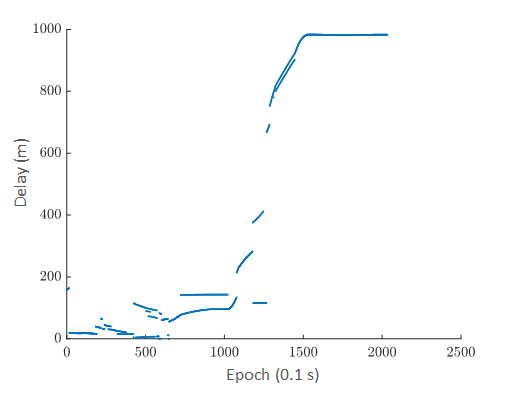}}
\caption{Multipath delay estimation for satellites generating NLOS}
\label{delay}
\end{figure}

Fig. \ref{delay} shows the multipath delays of the satellites generating NLOS estimated by the ray-tracing method (section III.C), where the x-axis represents the epoch and the y-axis represents the reflection delays of the satellites.
In addition, we compare the time spent on multipath estimation between using the proposed camera 3D reconstruction on line and using the existing offline commercial or open-source 3D building models, i.e. simulation map. The time spent on our accelerated reconstruction map multipath estimation is 2 minutes and the time for the pre-constructed simulation map is 22 minutes. Obviously, the proposed method accelerates the multipath estimation by almost 10 times. An important observation taken advantage of by the proposed method is that in the real environment, multipath does not occur at lower heights due to shading by trees in parts of the roadway, thus reducing the number of planes responsible for multipath. This height is not a fixed value and needs to be estimated according to the actual situation and scenario in an online manner, so the tree-shaded segments cannot be removed directly by presetting the height range of the aforementioned offline simulation map. The proposed method can reduce the number of planes and improve the computational efficiency of multipath estimation more effectively than the offline multipath estimation method.

\begin{figure}
\centerline{\includegraphics[width=18.5pc]{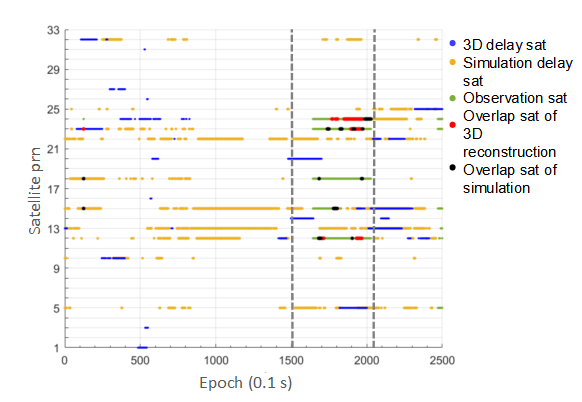}}
\caption{NLOS satellites and observed satellites comparison}
\label{NLOS compare}
\end{figure}

\begin{figure}
\centerline{\includegraphics[width=18.5pc]{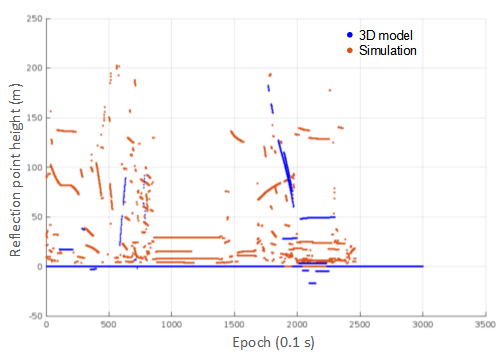}}
\caption{Reflection height comparison}
\label{Height}
\end{figure}

\begin{figure}
\centerline{\includegraphics[width=18.5pc]{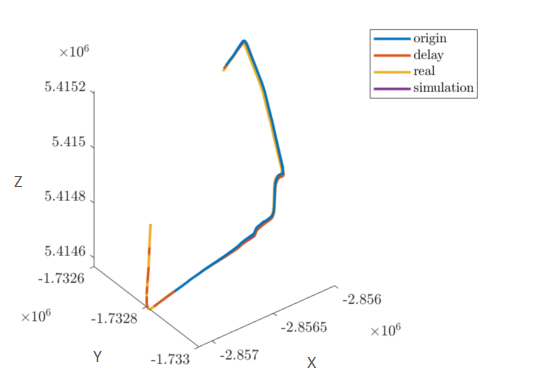}}
\caption{Comparison between positioning results}
\label{position result}
\end{figure}

\begin{figure*}
\centerline{\includegraphics[width=40.5pc]{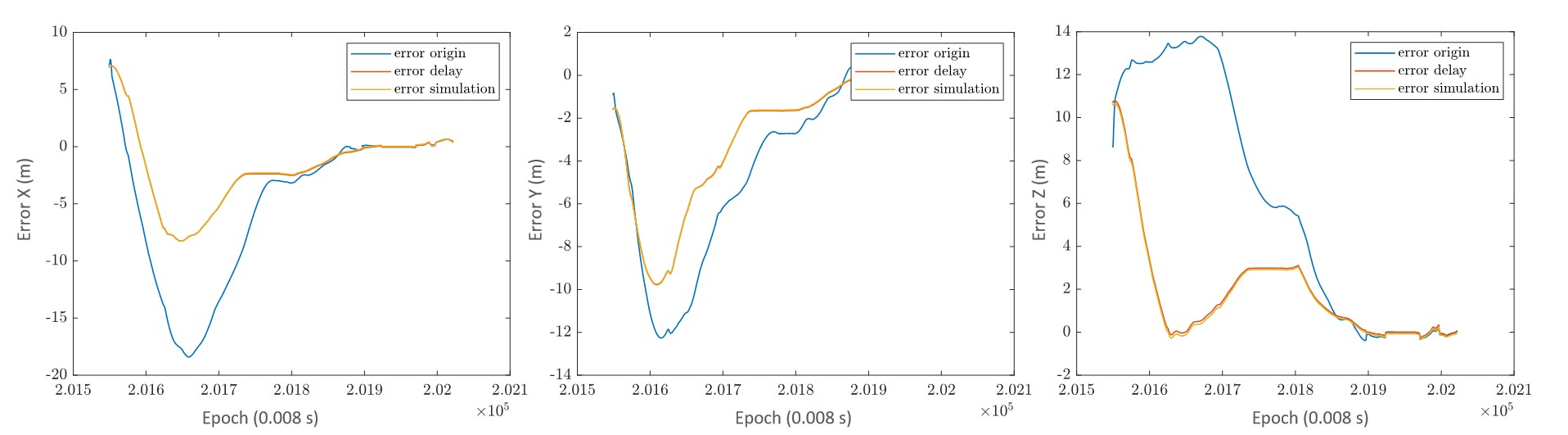}}
\caption{Error comparison in ECEF frame}
\label{error_XYZ}
\end{figure*}

We compared the number of NLOS satellites and the height of reflection points to do further analysis of the cause. Fig. \ref{NLOS compare} presents a comparative analysis of observed satellite PRNs (green dots), NLOS-delayed satellites identified through offline simulation map geometric analysis (yellow dots), and NLOS-delayed satellites detected via this work’s online 3D reconstruction-based geometric analysis (blue dots). The red dots represent the overlap of NLOS satellites indicated by this work's ray-tracing methods with the observed satellites indicated by a commercial receiver; the black dots represent the overlap of simulation NLOS satellites with the observed satellites. Observations reveal incomplete overlap between the observed satellites and those identified through ray-tracing-based reflection analysis. It is worth noting that the simulation-based approach estimates a much higher number of satellites affected by NLOS compared to the online 3D ray-tracing approach. As can be seen in Fig. \ref{Height}, this difference mainly stems from the large number of building sections between 10-20 m height in the simulated environment, a small part of the height range identified in Section IV.A as a major contributor to the reflection of NLOS signals. However,
field observations made while the vehicle was in motion indicated that vegetation obstructions on both sides of the road largely blocked the signal at NLOS distances in this
altitude range.

The 3D modeling approach relies on camera-captured environmental data that excludes most of the unrecorded structural elements in the 10-20 m height zone, inherently avoiding this overestimation. This selective data capture greatly reduces computational redundancy in reflection path calculations.

Positioning result comparisons are further analyzed for NLOS delayed satellites between 1500 - 2100 epochs. The estimations of the 3D ray-tracing and simulation maps are not exactly overlapped, but their corresponding satellites are consistent with some temporal deviations. The consistency of the two methods in the major interference epochs proves the technical feasibility of the proposed online 3D ray tracing, i.e. 3DMA method in the practical application of urban NLOS multipath assessment.

Fig. \ref{position result} shows the raw positioning results without multipath mitigation in ECEF coordinate system, the positioning results after multipath mitigation and the ground truth generated by a fiber optics INS+GNSS suite. Fig. \ref{error_XYZ} shows the errors computed using trajectory ground truth from this fiber optic INS+GNSS suite and NovAtel’s Inertial Explorer without processing the multipath delay, positioning result of multipath estimation using simulation model and the positioning result of multipath estimation using the proposed 3DMA method. Comparing the error plots, we can find the significant improvement in positioning accuracy after multipath estimation processing. This can verify the effectiveness of this method. However, the proposed method possesses a 10 times faster processing speed (2 min vs 20 min for a 4-min long data), which translates into 0.033 seconds per frame or 30 fps (frames per second).  Comparing the proposed method with the simulation method, our approach provides a better balance of computational efficiency and accuracy.

\subsection{Discussion}
The results validate the effectiveness of the proposed method. The orthogonal feature fusion framework and arbitrary-shaped boundary recovery method can effectively enhance the reconstruction granularity. The 3D model of the building façade generated based on the point cloud multiplanar fitting algorithm can accurately characterize the geometric features of the building surface. Compared with ray-tracing using pre-constructed offline (simulation) methods, the computational efficiency of multipath estimation is significantly improved by the proposed method while maintaining similar multipath delay corrective capability. This performance advantage stems from three improvements: firstly, the proposed orthogonal feature fusion 3D reconstruction model innovatively integrates the semantic sensing capabilities of SuperPoint with the geometric feature extraction strengths of the SIFT algorithm. This ensures accurate reconstruction of both texture-rich and texture-poor regions. Furthermore, this approach facilitates dynamic alignment between the 3D reconstruction area and the effective signal coverage area of the GNSS receiver, thereby preventing redundant modeling of non-covered areas at the source.  Secondly, the proposed refinement of point cloud segmentation to recover arbitrarily shaped building boundaries reduces the number of planes used for multipath estimation while maintaining topological consistency. Thirdly, a reflection point height range estimation method for GNSS NLOS signals is constructed to filter the effective reflection planes through the height \textit{a priori} information, which reduces the number of redundant planes while guaranteeing the multipath signal coverage, and thus improves the computational efficiency of multipath estimation.

Experiments show that the method can realize arbitrary-shaped high-precision surface reconstruction in strongly reflective glass curtain wall building complexes and thus effectively solve the problems such as specular reflection path mis-matching. In modern urban canyons, such complex shapes, for example with surface normal nonparallel to the ground surface, appear at a greater chance. Thus, the proposed method is expected to improve the multipath estimation accuracy by a decent margin.

However, the study still has some limitations. Firstly, the model simplification assumption ignores low-height dynamic objects and architectural detail features, which may lead to local multipath estimation bias. Secondly, there is a threshold constraint on the effective modeling distance due to the limitation of camera resolution. Thirdly, extreme weather conditions may affect the stability of visual reconstruction. Subsequent research will focus on fusing multimodal sensory data and developing adaptive algorithms to extend environmental adaptability.

To summarize, experimental results show that the method enables multipath estimation to achieve an optimal balance between positioning accuracy and computational efficiency, providing a new idea for real-time positioning in complex urban environments.

\section{CONCLUSION}
This study proposes an accelerated camera-only 3DMA framework for efficient urban GNSS multipath estimation and mitigation. By proposing an orthogonal feature fusion framework, we successfully handle the challenge of reconstruction robustness imposed by highly reflective surfaces; such surfaces are the major source of radio frequency signal reflection but traditionally could hardly be effectively reconstructed using camera. By establishing a polygon- instead of triangle- or rectangle-oriented surface reconstruction pipeline, we address the challenge of reconstruction granularity or accuracy, thus providing more convincing evidence of whether and where the GNSS signals \textit{are} actually reflected. Our approach further employs a multi-plane fitting algorithm to reduce the complexity of the 3D point cloud maps and propose two complexity control strategies to help reduce the processing burden of 3DMA ray-tracing, thus removing the obstacle for a new balance between computational efficiency and precision in GNSS multipath estimation. In addition, we conducted a real world data acquisition campaign in Lujiazui, Shanghai, a typical multipath-prone urban canyon, to verify the feasibility of the proposed algorithm. The results show that the method achieves an average reconstruction accuracy of 2.4 meters during a 4 minute ride in dense urban environments featuring glass curtain wall structures, a traditionally tough case for reconstruction, and achieves a ray-tracing-based multipath correction rate of 30 image frames per second, 10 times faster than the contemporary benchmarks.

\section*{ACKNOWLEDGMENT}
This work was supported by the National Key R$\&$D Program of China (2022YFB3904401).

\section*{REFERENCES AND FOOTNOTES}

\end{document}